\documentclass[12pt,preprint]{aastex}

\newcommand{\rC}{{$^{12}$C/$^{13}$C}{}}
\newcommand{\rN}{{$^{14}$N/$^{15}$N}{}}

\shorttitle{DEEP IMPACT WITH UVES AT VLT AND HIRES AT KECK 1}
\shortauthors{JEHIN, MANFROID, HUTSEM\'EKERS ET AL.}

\begin{document}


\title{Deep Impact : High Resolution Optical Spectroscopy \\
with the ESO VLT and the Keck 1 telescope}


\author{
E. Jehin\altaffilmark{1},
J. Manfroid\altaffilmark{2},
D. Hutsem\'ekers\altaffilmark{2},
A.L. Cochran\altaffilmark{3},
C. Arpigny\altaffilmark{2},
W. M. Jackson\altaffilmark{4},
H. Rauer\altaffilmark{5},
R. Schulz\altaffilmark{6} and
J.-M. Zucconi\altaffilmark{7}
}

\altaffiltext{1}{European Southern Observatory, Casilla 19001, Santiago, 
Chile; ejehin@eso.org}
\altaffiltext{2}{Institut d'Astrophysique et de G\'eophysique, Sart-Tilman, B-4000 Li\`ege, Belgium; manfroid@astro.ulg.ac.be; arpigny@astro.ulg.ac.be; hutsemekers@astro.ulg.ac.be}
\altaffiltext{3}{Department of Astronomy and McDonald Observatory, University 
of Texas at Austin, C-1400, Austin, USA; anita@barolo.as.utexas.edu}
\altaffiltext{4}{Department of Chemistry, University of California, 1 Shields Avenue, Davis, CA 95616; wmjackson@ucdavis.edu}
\altaffiltext{5}{Institute of Planetary Research, DLR, Rutherfordstr. 2, 12489 Berlin, Germany; Heike.Rauer@dlr.de}
\altaffiltext{6}{ESA/RSSD, ESTEC, P.O. Box 299, NL-2200 AG Noordwijk, 
The Netherlands; rschulz@rssd.esa.int}
\altaffiltext{7}{Observatoire de Besan\c{c}on, F25010 Besan\c{c}on Cedex, 
France; jmz@dalai-zebu.org}

\begin{abstract}

We report on observations of comet 9P/Tempel\,1 carried out before, during,
and after the NASA DEEP IMPACT event (UT July 4), with the optical 
spectrometers UVES and HIRES mounted on the telescopes Kueyen of the 
ESO VLT (Chile) and Keck 1 on Mauna Kea (Hawaii), respectively. A total 
observing time of 
about 60 hours, distributed over 15 nights around the impact date, allowed us 
(i) to find a periodic variation of 1.709 $\pm$ 0.009 day in the CN and NH 
flux, explained by the presence of two major active regions;
(ii) to derive a lifetime $\gtrsim$ 5 x 10$^4$ s for the parent of the CN 
radical from a simple modeling of the CN light curve after the impact; 
(iii) to follow the gas and dust spatial profiles evolution during the 4 
hours following the impact and derive the projected velocities (400 m/s
and 150 m/s respectively); 
(iv) to show that the material released by the impact has the same carbon 
and nitrogen isotopic composition as the surface material 
(\rC\ = $95\pm15$ and \rN\ = $145\pm20$).

\end{abstract}


\keywords{Comets : general --- comets: individual (9P/Tempel 1) 
--- Solar System: formation}

%\clearpage
\section{OBSERVATIONS}

High-resolution spectra of comet Tempel\,1, the target of the NASA DEEP 
IMPACT (DI) mission, have been collected with the UV-Visual Echelle 
Spectrograph (UVES), of the ESO VLT in early June (UT 2, 7 and 8) and during 
a 10 night run around the DI event (UT July 2-11). The nights were 
photometric or clear and the seeing excellent to good (0.4"-1.1"). Two long 
exposures (of $\sim$7200s) have been secured each night using two different 
beam splitters, a combination which allowed us to obtain on each 
observing date a spectrum covering the full optical range (304-1040nm, except
 for a few narrow gaps). Both settings were chosen in order to include the CN
 (0-0) Violet band near 388\,nm. The narrow slit of the spectrograph (0.44" x
 10" or 288\,km x 6540\,km on July 4) provides a resolving power 
$\lambda/\Delta\lambda$ $\sim$ 83,000 and was generally put on the center of 
light of the comet (and in a few cases at about 1.0" from it). The position 
angle of the slit was along or perpendicular to the Sun-comet vector most of 
the time and an atmospheric dispersion corrector (ADC) was used to sample the
 same region during a given exposure.

On July 4, the comet was setting at Paranal Observatory at the exact time of 
the DI event UT 05:52 \citep{AHearn05}. The acquisition of the last UVES 
spectrum ended at UT 02:59 which was just a couple of hours before the impact, 
i.e., before the first spectrum obtained with HIRES, the High 
Resolution Echelle Spectrometer of the Keck 1 telescope installed 
on Mauna Kea (Hawaii). The comet was visible again from Paranal 17 hours 
later. The data sets are complementary: the UVES one nicely describes the 
pre- and post- impact behaviour of the comet, while HIRES data contain unique
 information about the direct consequences of the impact (up to 4 hours 
later).

The spectra have been reduced with special emphasis on the orders showing the
 CN band at 388\,nm. We used the echelle package of the IRAF software (NOAO) 
to calibrate and extract the spectra. The dust-reflected sunlight underlying 
the cometary emissions was removed by subtracting a solar reference spectrum 
(the Moon, in this case) after the appropriate Doppler shift, profile fitting
 and normalization were applied.

\section{CN PERIODIC VARIATIONS}

The relative flux in the CN band - between 386.2 and 387.5\,nm and 
integrated over the full slit area - was calculated for the 13 nights of 
UVES observations. 

The goal of the flux measurements was first to evaluate the variations triggered 
by the impact. 
But obvious flux variations of up to 30\% were visible from night to night. 
A period search analysis has been performed using Fourier series fits and other 
methods like the Renson algorithm \citep{Ren78} better suited for anharmonic light 
curves. It yielded a period of 1.709 $\pm$ 0.009 day (40.86 $\pm$ 0.05 hrs) in very 
good agreement with the 1.701 $\pm$ 0.014 day rotation period determined from the 
DI spacecraft nucleus lightcurve \citep{AHearn05}.  

The phase diagram of the CN flux using this period is shown in Fig.1. 
The flux modulation is obviously synchronized with the rotation period and was stable 
 over more than one month. The only deviating points are those corresponding to the 
first four post-impact measurements obtained on UT July 04 and 05 (phases around 0.45 and 0.1). 
 There is no evidence of strong sporadic activity in the data. None of the 13 UVES 
observing dates corresponds to the outbursts detected by the DI spacecraft or the 
ground based observatories.
After removing the periodic background, the light curve shows a net flux excess of about 
20\% and 7\% respectively, 17 and 41 hours after the impact, indicating that the gas released 
by the ejected material had not yet completely vanished. After July 7.0 
(and up to July 12.0), no excess could be detected at all in the region studied. 

The shape of the CN phase diagram may be explained by the periodic passage of two 
major active regions, as well as a weaker one, into sunlight.
Such an interpretation was also used to explain the comet Halley photometric observations 
during its 1985/1986 apparition \citep{Sch90}. 
The brightest feature starts to produce gas around phase 0.9 (its minimum) and reaches 
its maximum (when the source stops its production) at phase 0.1. The second region 
is $\sim$20\% less active and lasts also for about 8 hours (from phase 0.6 to 0.8). 
They could both be located at the same latitude (they see the Sun during the same 
amount of time), the difference in activity resulting from a different size or 
sunlight illumination of the active regions due for instance
to different land morphology. The distance between the two features
is a bit more than 1/3 of the rotation. The shoulder at phase $\sim$ 0.4 could be some 
evidence for the presence of a third and much fainter source.

The impact occurred at the beginning of the strongest periodic 
brightening (Fig.1). It is then possible that this active region is visible close to 
the terminator in the images sent by the impactor. As already noted by \citet{AHearn05}, 
a possible candidate is the large and smooth area (labeled ``a'') in their Fig.1
and the scarp to the north of this feature would be an excellent candidate for a more 
localised, fast reacting region (the large smooth area being a kind of reservoir).  

It is highly significant to note that almost all outbursts observed by the DI spacecraft
(we have a precise knowledge of their onset time \citep{AHearn05}) correspond to minima 
in our gas phase diagrams (see Fig.1). Moreover 4 of the 6 outbursts are located just before the 
strongest brightening (phases from 0.85 to 0.95). Thus, the outbursts seem clearly
associated with the active regions we found (they occur at the same phases) and would be 
occasionally triggered when one of these regions comes into daylight. The outbursts could 
just be a particularly strong phase of outgassing from the same region. 

The flux in the NH 0-0 (A-X) band at 336 nm shows a nice correlation with the CN variations 
and any phase shift must be small, at most a couple of hours (Fig.1). The first post-impact 
measurement has an excess of 30\%, i.e., significantly more than CN, but, contrary to CN, the second 
one is not enhanced (or only slightly), which means that NH disapeared sooner
after the impact. Many more species are available in our spectra (OH, C$_3$, C$_2$, NH$_2$, etc.)
and will be examined in the same way. This may provide interesting information on
their respective parent lifetimes. For instance, the NH parent seems to have a similar
lifetime to the CN parent lifetime because the phase-shift between the two species is very small.
With a lifetime in the range 2-6 10$^4$ s \citep{Wyc88, Fin91} for NH$_2$, 
similar to that of the CN parent (2-5 10$^4$ s, \citet{Fra05} and references therein), this 
observation is indeed in agreement with NH$_2$ being the main NH parent \citep{Fin91}.

Data on HCN - a possible parent candidate of CN - have been obtained with the IRAM 30-m 
radio telescope from May 4.8 to 9.0 and show a 1.67$\pm$0.07 day periodicity \citep{Biv05}. 
The beam used sampled similar regions to UVES ($\sim$10").
The data have been phased with the 1.709 day period and superposed with the UVES CN flux 
measurements. 
There are only five HCN measurements, with large errors, so it is not possible
to verify if the differences we see in shape and phase with our CN lightcurve are real.
If this is the case, this could be some additional evidence that HCN is not the only
parent of CN \citep{Fra05, Man05}.

\section{CN AND DUST IMPACT LIGHTCURVES}

Three pre-impact spectra were obtained on May 30 with the HIRES spectrograph at the 
Keck 1 telescope and a series of 14 relatively short exposures (10 to 30 minutes), with 
one right before the impact, were taken on the DI night\footnote{Those data are 
publicly available at http://msc.caltech.edu/deepimpact/}. 
The 0.86'' by 7'' slit provides a resolving power $\lambda/\Delta\lambda$ $\sim$ 48,000 
and samples a comparable zone (563 km x 4578 km) 
to that of UVES. The slit was always centered on the nucleus and aligned along 
the parallactic angle. The weather was photometric on July 4 and the seeing 
excellent ($\sim$0.6'').

The total CN flux in the slit has been calculated and normalized 
to the UVES fluxes by comparing the first, pre-impact spectrum of July 4, as well 
as the three May spectra with our derived lightcurve. 
The HIRES slit was 2x wider than the UVES one and always centered on the nucleus,
leading to better measurement of the dust component. The consequences of the impact 
are readily seen in Fig.2, both in the dust and CN flux. The CN periodic
variation has been removed using the UVES lightcurve and the dust background 
has been taken out using the pre-impact spectrum. 

The total CN flux reaches its maximum at UT 07:28 ($\pm$ 10min) or 5760 $\pm$ 600 seconds 
after the impact. It is enhanced by a factor of 2.8 compared with the pre-impact situation.
The decline is explained (at least partially) by the molecules starting to leave the slit
area, mostly through the slit length (the lifetime of the CN parent is indeed long 
enough \citep{Rau03} to make the escape from the slit width unnoticed).
The average projected speed of CN to cross the half slit is then $\sim$ 0.40 $\pm$ 
0.04 km/s. This is smaller than typical molecular outflow velocities in 
cometary comae \citep{Com05} but it is only a lower limit due to the 
projection effect. 
The CN peak intensity of the spatial profiles is reached well before the total CN flux
maximum, at about UT 06:30 ($\pm$ 10min).  

The CN light curve contains valuable information about the dissociation lifetime of
 its parent molecule but a complete modeling is beyond the scope of this paper.
To first approximation, the CN impact light curve may be interpreted by assuming 
a competition between the CN creation resulting from the dissociation of a parent 
molecule ($\propto (1-\exp(-t/\tau))$ where $\tau$ is the parent lifetime) and the 
exit from the slit area ($\propto t^{-1}$ after filling the slit width and 
$t^{-2}$ after filling the slit length).
A precise determination of the parent lifetime from our data would require a good
knowledge of the CN velocity distribution. The radial profiles (Fig.3) indicate a range of 
projected velocities from 250 to 650 m/s, in good agreement with the mean velocity 
of 400 m/s deduced from the light curve. Adopting 400 m/s, the slow decline of the light
curve indicates that CN creation still occurs after filling the slit area, requiring 
relatively long parent lifetimes in agreement with those determined using the Rosetta 
spacecraft data ($\tau$ $\sim$50000 s) \citep{Kel05}.

The excess of the CN flux observed with UVES until two days after the impact
 indicates that the decline might have occured in two phases, a fast one documented by 
HIRES, and a slow one lasting over more than a full rotation of the comet. 
A linear extrapolation of the flux decrease observed by HIRES after the impact shows 
that the extra CN should have been gone from the studied area about 8 hours after the impact. 
The excess measured in UVES data could be explained by remnant activity of the 
crater or the presence of another minor CN parent with a longer lifetime.  
Unfortunately, data are missing to describe the lightcurve between MJD 53555.4 and 53556.0. 

Taking the observed time evolution of the CN excess during 
the impact as the response of the comet to an elementary event, we tried to model the 
out-of-impact CN lightcurve as the response to a series of such elementary pulses. 
The observations can be accurately reproduced with just three bursts (or jets, corresponding 
to the observed bumps) (see Fig.1). These are simulated by convolving the elementary response 
with a simple temporal variation representing the instantaneous production of each source.
There is no room for a ``background" CN flux, i.e., all CN is produced by these periodic 
jets starting to release gas at phases 0.25, 0.45 and 0.85 (intensity of, respectively, 
0.2, 0.7 and 1.0).
The full width at half maximum duration of the bursts is 8 hours assuming a 
quasi-symmetrical triangular shape to the intensity of the jets. This behaviour equally 
explains NH and presumably all gases. 

The variations of the dust after the impact are much faster, with a very steep brightening. 
The maximum is already reached  at about 
UT 06:18 $\pm$ 10\,min, the dust being at that time enhanced by a factor $\sim$ 8.5 with 
respect to the pre-impact spectrum. Contrary to CN, the intensity peak in the dust profiles 
is reached at about the same time as the spatially integrated flux.
The decline of the dust emission may also be interpreted by the escape of dust from 
the slit. This would give a projected dust velocity of $\sim$ 0.18 $\pm$ 0.05 km/s,
as, in this case, the escape from the slit width will be the dominant factor.  
However the slow and quasi linear decline would require a rather broad range of velocities 
-- slower than the gas component (Fig.3) -- and/or complex processes like destruction 
of highly reflective icy grains by sunlight.
The radial profiles show that the dust is expanding at a much slower pace 
than CN, at about 0.13 $\pm$ 0.03 km/s during the first 1.5 hour after the impact. 
This is consistent with the value given above and what others have measured \citep{Kel05,
Sug05}.

There is a slight but clear increasing shift of the position of the CN intensity peak in
the radial profiles with respect to the dust (nucleus dominated) peak (Fig.3). The shift
of $\sim$350\,km (at 07:10 UT, and at a mean PA=215$^\circ$) is in the direction of the
plume ejecta and is most certainly associated with it. It may indicate a preferential ejection 
direction for the gas with respect to the dust, the gas being emitted towards the Sun.

\section{ISOTOPIC RATIOS}

Measuring the isotopic ratios in 9P/Tempel\,1 before and after the impact was a unique 
opportunity to check whether the material buried several meters below the surface and released
by the impact is different or not from what we usually observe.

It was important to obtain a high quality pre-impact spectrum in order to compare it
with post-impact data. This was achieved with UVES thanks to the 10 hours obtained 
in June and the 8 hours during the two pre-impact nights. The individual CN (0-0) Violet 
band spectra were combined after extraction with an optimal weighting scheme in 
order to maximize the overall signal-to-noise ratio. Synthetic spectra
of the different CN isotopes were computed for each observing circumstance following 
the scheme described by Zucconi and Festou \citep{Zuc85}. The isotope mixture was then 
adjusted to best fit the observed final spectrum 
\citep{Arp03, Jeh04}. The same was done for the UVES post-impact data, and as no difference 
was found (which is not too surprising as those spectra were only slightly affected 
by the impact (Fig.1)), all the spectra were combined to produce a single 50 hours spectrum.
The best fit is obtained for an isotopic mixture \rC$=(95\pm15)$ and \rN$=(145\pm20)$.

From the HIRES impact lightcurve (Fig.2) we extracted the spectra showing a brightening 
of more than a factor two (the 10 spectra from 06:33 to 09:11 UT). Thus in those spectra 
half the CN flux might come from the fresh material released by the impactor. 
Despite the shorter exposure time (about 4 hours of total exposure time) and the
lower resolving power of the HIRES spectrum compared to the UVES out of impact spectrum, 
the large collecting area of the Keck 1 telescope and the relative brightening of the 
comet allowed us to determine the following values \rC$=(95\pm15)$ and \rN$=(165\pm30)$ 
for the carbon and nitrogen isotopic ratios.
Those values are compatible with the values determined out of impact. It appears that 
the \rN\ ratio of the burried material is still most probably below the solar value \citep{Arp03}.

After comet 88P/Howell, comet 9P/Tempel\,1 is the second Jupiter-family 
comet to have a \rN\ ratio determination. Both values are in excellent agreement and are
similar to the ratios measured in half a dozen Oort-cloud comets \citep{Hut05}. 
The fact that the ejected material - supposed to be pristine as it was never exposed to 
the solar radiation and cosmic rays - has the same isotopic composition, 
favors an isotopic homogeneity between the two populations of comets, despite the fact 
that they are expected to have formed at very different locations in the solar system \citep{Wei99}. 
This is a strong argument in favor of a primordial origin of the high content of
$^{15}$N in cometary volatiles. This peculiarity is not the result of some alteration process 
of the comet's surface material and was already present in the protosolar nebula 
before the accretion process, which gave birth to the comets and planets, took place. 
In case the N isotopic ratio changes within the solar nebula, this is a new argument in 
favor of the volatile ices in Jupiter-family and Oort-cloud comets 
originating in a common region of the protoplanetary disk \citep{Mum05}.  
 
\acknowledgments
Based on observations carried out at the European Southern Observatory (ESO) 
under prog. 075.C-0355(A). This program was part of a joint European initiative in 
support of the NASA DEEP IMPACT mission to comet 9P/Tempel\,1 by ground-based 
observations at La Silla and VLT Paranal. 
Some of the data presented herein were obtained at the W.M. Keck Observatory, 
which is operated as a scientific partnership among the California Institute 
of Technology, the University of California and the National Aeronautics and 
Space Administration. The Observatory was made possible by the generous 
financial support of the W.M. Keck Foundation.
JM is Research Director and DH is Research Associate at FNRS (Belgium).

%\clearpage

%\clearpage

%---------------------------------------------------------------------
\begin{figure}
\plotone{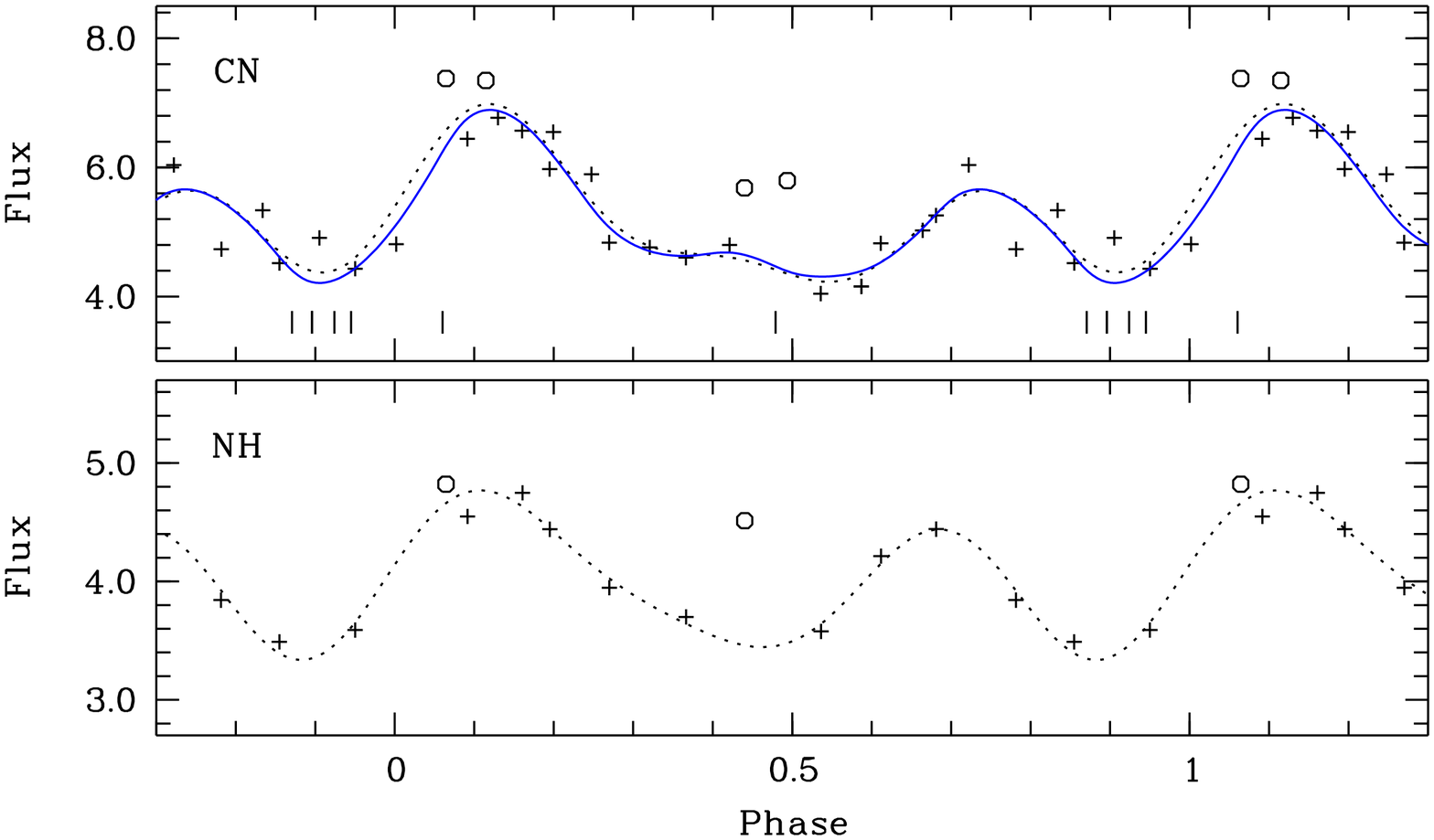}
\caption{
The UVES CN and NH gas emission rotational phase diagrams. Zero phase is at the time of 
impact as observed from Earth : MJD 53555.2445 (July 4, UT 05:52:02). 
In this figure, and in the following ones, the  fluxes are given in arbitrary units.
The dashed line models of the gas "lightcurves" are based on a three harmonic least
squares fit in which each data point has the same weight. 
The fit of the NH data is independent of the CN one and the similarities of the two curves 
should be noted.
Both data sets are phased to P = 1.709 day and reveal the existence of two main active 
regions.
The observations can be accurately reproduced (continous line) with just three periodic bursts 
(or jets, corresponding to the observed bumps) starting to release gas at phases 0.25, 0.45 
and 0.85 and with a duration of 8 hours (see text).
The enhanced post impact data for UT July 4 and 5 (around 23:00) are marked as circles and
correspond to phases 0.44 and 0.49 and 0.06 and 0.11, respectively. 
The phase of the outbursts detected by the DI spacecraft are indicated as tick marks
at the bottom of the figure. 
The scatter in the data comes from the centering and orientation of the slit, 
sky transparency or intrinsic variability of the sources. The NH 0-0 (A-X) emissions at 336 nm 
 appear only in the bluest setup so we have only one data point per night for that species.
}
\end{figure}
% ------------------------------------------------------------------

%\clearpage

%---------------------------------------------------------------------
\begin{figure}
\plotone{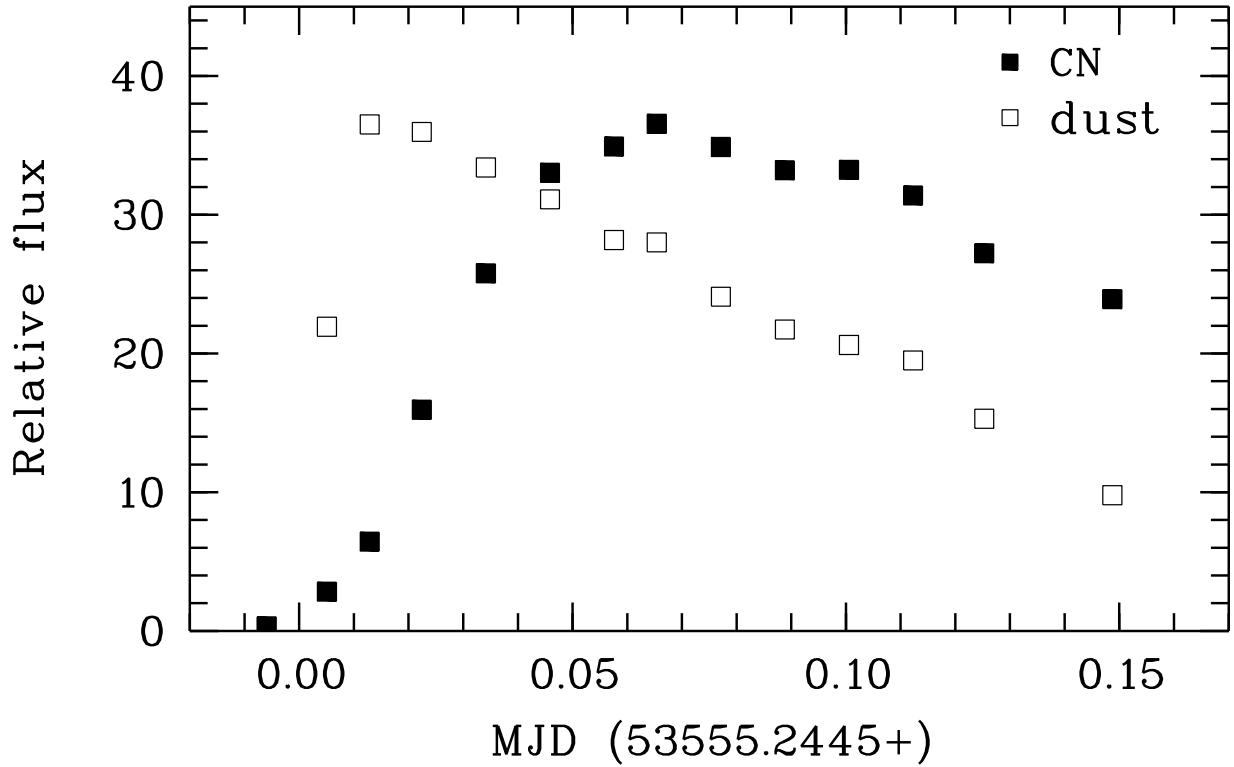}
\caption{
The HIRES CN and dust impact lightcurves.
Time 0.0 is the time of the impact as observed from Earth : MJD 53555.2445 
(July 4, UT 05:52:02). 
Each data point represents for each spectrum (at mid exposure) the 
total (spatially integrated) flux in the CN (0-0) violet band (388\,nm) and in the dust 
continuum underlying the CN features (empty and filled squares respectively). 
The fluxes are normalised to the exposure time, corrected for airmass, and the respective 
backgrounds are subtracted. The CN curve is scaled (x3.5) to the dust curve for 
easier comparaison. 
}
\end{figure}
% ------------------------------------------------------------------

%\clearpage

%---------------------------------------------------------------------
\begin{figure}
\plotone{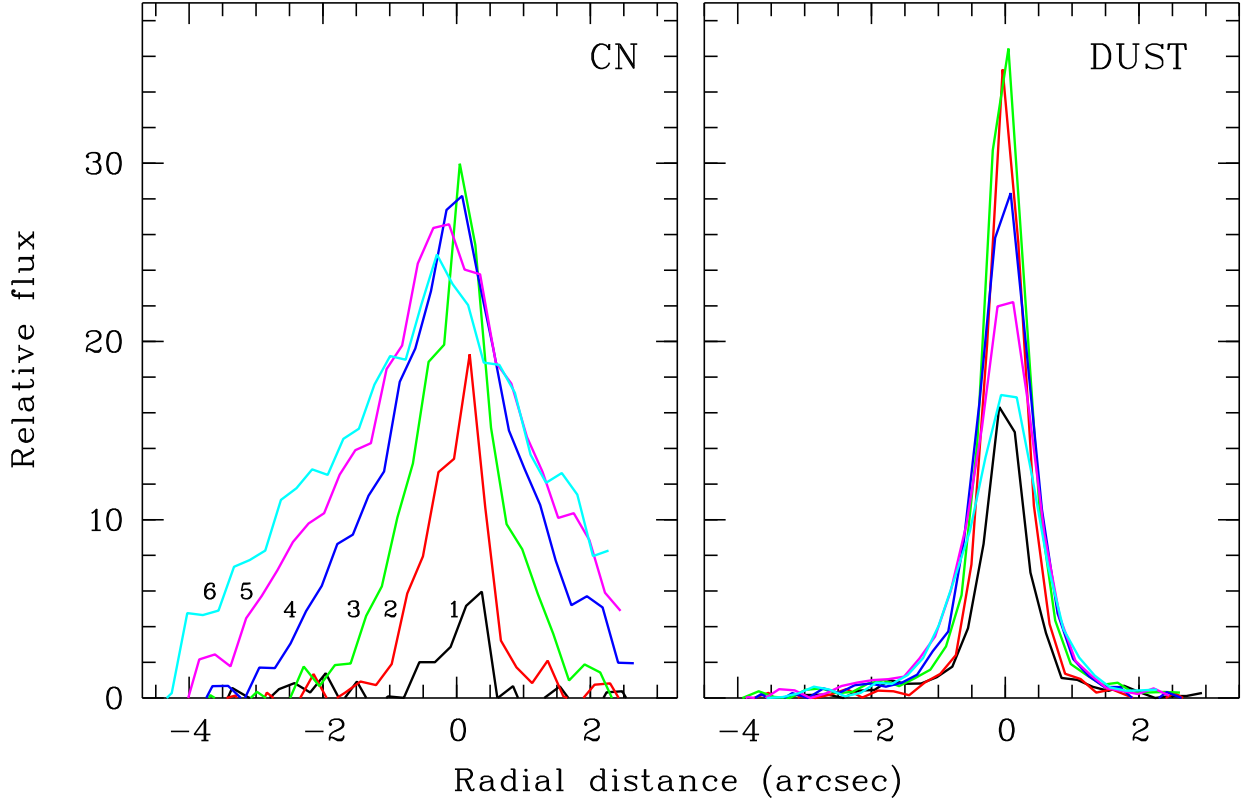}
\caption{
The HIRES CN and dust radial profiles during the first hour following the impact.
The profiles of the first 6 post-impact spectra (respectively at mid exposure time
06:00:18.0, 06:11:12.0, 06:24:35.0, 06:40:29.0, 06:56:22.0 and 07:12:16.0 UT, July 4) 
are extracted along the 7'' slit of the spectrograph and background subtracted.
The pixel scale in spatial direction is 0.239'' or 155 km. 
The slit was set along the parallactic angle with the consequence that the position angle (PA) 
on sky was changing during the observations (20.1$^{\circ}$, 24.8$^{\circ}$, 29.7$^{\circ}$, 
35.5$^{\circ}$, 40.6$^{\circ}$ and 48.9$^{\circ}$ for, respectively, the first 6 post-impact spectra).
%The slit was set along the parallactic 
%angle and changing from position angle PA=20$^{\circ}$ to 50$^{\circ}$ during the observations. 
This direction is close to the axis of the expanding cloud (PA=225$^{\circ}$, \citet{AHearn05}). 
Note that the PA of the extended Sun-comet radius vector is 111$^{\circ}$.
The profiles are recentered on the dust peak to compensate for the shift along the
slit induced by the atmospheric refraction. The different behaviour
of the gas and the dust is obvious.
}
\end{figure}
% ------------------------------------------------------------------

\end{document}